\documentclass[aps,prb,superscriptaddress,twocolumn,reprint,amsmath,amssymb]{revtex4-2}
\usepackage{graphicx,color}

\begin{document}

\title{Superconductivity of highly spin-polarized electrons in FeSe probed by $^{77}$Se NMR}

\author{S. Molatta}
\author{D. Opherden}
\author{J. Wosnitza}
\affiliation{Hochfeld-Magnetlabor Dresden (HLD-EMFL) and W\"urzburg-Dresden
Cluster of Excellence ct.qmat, Helmholtz-Zentrum Dresden-Rossendorf,
01328 Dresden, Germany}
\affiliation{Institut f\"ur Festk\"orper- und Materialphysik, TU Dresden,
01062 Dresden Germany}
\author{Z. T. Zhang}
\affiliation{Hochfeld-Magnetlabor Dresden (HLD-EMFL) and W\"urzburg-Dresden
Cluster of Excellence ct.qmat, Helmholtz-Zentrum Dresden-Rossendorf,
01328 Dresden, Germany}
\affiliation{Anhui Province Key Laboratory of Condensed Matter Physics at Extreme Conditions, High Magnetic Field Laboratory, Chinese Academy of Sciences, Hefei 230031, China}

\author{T. Wolf}
\affiliation{Institut f\"ur Quantenmaterialien und -technologien, Karlsruher Institut f\"ur Technologie, 76021 Karlsruhe, Germany}

\author{H. v. L\"ohneysen}
\affiliation{Institut f\"ur Quantenmaterialien und -technologien, Karlsruher Institut
f\"ur Technologie, 76021 Karlsruhe, Germany}
\affiliation{Physikalisches Institut, Karlsruher Institut f\"ur Technologie, 76049 Karlsruhe, Germany}

\author{R. Sarkar}
\affiliation{Institut f\"ur Festk\"orper- und Materialphysik, TU Dresden,
01062 Dresden Germany}

\author{P. K. Biswas}
\affiliation{ISIS Facility, Rutherford Appleton Laboratory, Chilton,
Didcot Oxon, OX11 0QX, United Kingdom}

\author{H.-J. Grafe}
\affiliation{IFW Dresden, Institute for Solid State Research,
01171 Dresden, Germany}

\author{H. K\"uhne}
\email[Corresponding author. E-mail: ]{h.kuehne@hzdr.de}
\affiliation{Hochfeld-Magnetlabor Dresden (HLD-EMFL) and W\"urzburg-Dresden
Cluster of Excellence ct.qmat, Helmholtz-Zentrum Dresden-Rossendorf,
01328 Dresden, Germany}

\date{\today}

\begin{abstract}
	
A number of recent experiments indicate that the iron-chalcogenide FeSe
provides the long-sought possibility to study bulk superconductivity in
the cross-over regime between the weakly coupled Bardeen--Cooper--Schrieffer
(BCS) pairing and the strongly coupled Bose--Einstein condensation (BEC).
We report on $^{77}$Se nuclear magnetic resonance experiments of FeSe,
focused on the superconducting phase for strong magnetic fields applied
along the $c$ axis, where a distinct state with large spin polarization
was reported. We determine this high-field state as bulk superconducting
with high spatial homogeneity of the low-energy spin fluctuations.
Further, we find that the static spin susceptibility becomes unusually
small at temperatures approaching the superconducting state, despite the
presence of pronounced spin fluctuations. Taken together, our results
clearly indicate that FeSe indeed features an unusual field-induced
superconducting state of a highly spin-polarized Fermi liquid in the
BCS-BEC crossover regime.
\end{abstract}
\maketitle

\section{Introduction}	
The discovery of superconductivity in iron-based materials in 2008 opened
a new avenue for the exploration of unusual superconducting states and
phenomena in multiband superconductors, with wide possibilities for
tuning the electronic ground states by variation of the material composition
and the ensuing thermodynamic parameters \cite{Kamihara2008,Johnston2010,Stewart2011,
Chubukov2015}. At present, the binary iron-chalcogenide FeSe attracts a lot
of research interest \cite{hsupnas105,Kasahara2016,bourgeoisPRL2016,baeknmat14,boehmerprl114,
wangnmat15,wangnphys11,ok20,Grinenko2018,Li2020,kas20}. As to the Fermi-surface topology
and energy scales in this material, there is one shallow hole pocket at
the $\Gamma$ point, and at least one electron pocket at the $M$ point,
both with remarkably small Fermi energies ($\varepsilon_F^e \approx$ 3 meV
and $\varepsilon_F^h\,\approx$ 10 meV, respectively) \cite{Kasahara2016,
Terashima2014}. The superconducting gap energies are $\Delta_1 =$ 2.5 meV
and $\Delta_2 =$ 3.5 meV, resulting in an unusually high ratio
$\Delta/\varepsilon_F \approx 1$ and $\approx 0.3$ for the electron and
hole bands, respectively \cite{Kasahara2016}.

Despite its structural simplicity, FeSe yields a wealth of interesting
phenomena. For example, a nematic transition, presumably induced by
orbital ordering, occurs at about 90 K, where the $C_4$ rotational
symmetry is broken,
while the translational symmetry is preserved \cite{baeknmat14,
boehmerprl114,wangnmat15,wangnphys11,boehmerarxiv150505120}. A drastic
increase of $T_c$ can be achieved via hydrostatic pressure, or when
growing thin films on specific substrates \cite{Medvedev2009,He2013,Ge2015}.

Further, very recently compelling evidence for the appearance of a
Fulde-Ferrell-Larkin-Ovchinnikov (FFLO) state was found when the
applied magnetic field is aligned parallel to the $ab$ plane \cite{kas20}.
On the other hand, for fields perpendicular to the planes, the observation
of an unusual superconducting phase with extremely high spin polarization,
dubbed B phase, was reported based on measurements of thermal transport
properties \cite{kasaharapnas111,watashigearxiv1607}. Close to the upper
critical field of superconductivity, all three relevant energy scales,
i.e., those of the Fermi energy, the superconducting gap, and the Zeeman
interaction, are of comparable magnitude, the combined action of which
may lead to a significant modification of the underlying electronic system.
For that reason, the condensation of electron pairs in the B phase is
proposed to take place in the BCS (Bardeen--Cooper--Schrieffer) -- BEC
(Bose--Einstein-condensate) crossover regime, which bridges the two
fundamental theories for the condensation of attractively coupled fermions.
In this crossover regime, the average interparticle distance approaches the size
of the interacting pairs, i.e., $k_F\xi\approx$ 1, where $k_F$ is the
Fermi wave vector and $\xi$ is the superconductive coherence length \cite{kasaharapnas111}. The resulting
ground state is a strongly interacting superfluid, out of which new
states of matter may emerge. The manifestation of preformed pairs with
an associated pseudogap, existing at temperatures much higher than
the actual condensation temperature, is a hallmark of the BCS-BEC crossover \cite{Randeria2014}.
As the intrinsic energy scales in materials usually place the electronic
interactions strictly in either the BCS or the BEC limit, little is
known about bulk superconductivity in the crossover regime.

In this paper, we present $^{77}$Se nuclear magnetic resonance (NMR)
data of a high-quality single crystal of FeSe. The parameter range of
our study covers magnetic fields between 5 and 16 T applied along the
crystallographic $c$ axis, and temperatures between 0.3 and 40 K.
Our work mainly focuses on the electronic properties at low temperatures
and close to the upper critical field of superconductivity, where the
existence of the B phase was reported, thus constituting the first investigation of the local dynamic properties across the A-B transition in FeSe at the BCS-BEC crossover regime.

\begin{figure}[tbp]
	\centering
	\includegraphics[width=0.95\columnwidth]{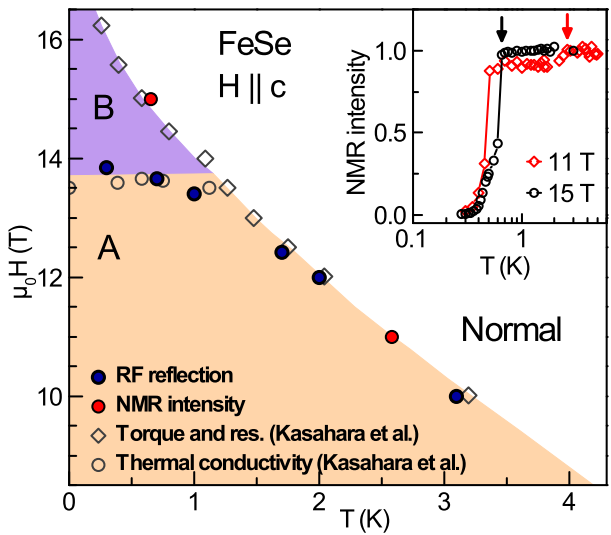}
	\caption{\label{fig:pd}
		Phase diagram of FeSe for fields parallel to the $c$ axis. RF reflection
		experiments (blue circles) probe superconducting transitions to the A phase,
		in excellent agreement with magnetic torque, resistivity, and
		thermal-conductivity results \cite{kasaharapnas111}, but are not sensitive
		to the onset of the B phase. Probing the $^{77}$Se NMR intensity (red
		circles), however, reveals the transition to the B phase---like the A
		phase---as a bulk superconducting state. Inset: Temperature dependence of the
		$^{77}$Se NMR signal intensity at 11 (red diamonds) and 15 T (black circles),
		the arrows indicate $T_c$.}
\end{figure}

\section{Experimental Methods}
The NMR coil with a vapor-grown FeSe single crystal \cite{boh13}, with
dimensions of approximately $5.0\,\times\,2.0\,\times\,0.04\,\text{mm}^3$,
was mounted on a single-axis rotator and placed directly in liquid
$^3$He inside a top-loading cryostat in a 16 T high-resolution magnet.
The $^{77}$Se NMR spectra were recorded using a commercial solid-state spectrometer and a standard Hahn spin-echo pulse sequence. The orientation of the magnetic field parallel to the crystallographic $c$-axis was adjusted with an
accuracy of about $\pm 2^{\circ}$ by probing the anisotropic frequency
shift of the $^{77}$Se spectra. To determine the superconductive phase
diagram of our sample, we also monitored changes of the complex radio-frequency
(RF) reflection at the NMR tank circuit. 

The nuclear spin-lattice relaxation rate is defined as $1/T_{1}T
\propto \sum_{q,n,m} F_{n m}(q)\chi_{nm}^{\prime\prime}(q,f_{res})/
f_{res}$, with $n,m = {x,y,z}$. Here, $F_{n m}$ denotes the hyperfine
form factors and $\chi_{n m}^{\prime\prime}$ is the imaginary part of
the dynamic electronic susceptibility. $T_1$ was measured via the
saturation-recovery method and determined by fitting $M_{z}(\tau)=
M_{0}\{1-\exp[-(\frac{\tau}{T_{1}})^{\beta}]\}$ to the recovery of the
nuclear magnetization after saturation, where $\beta$ is a stretching
exponent that accounts for a distribution of relaxation times.

\begin{figure}[tbp]
	\centering
	\includegraphics[width=0.95\columnwidth]{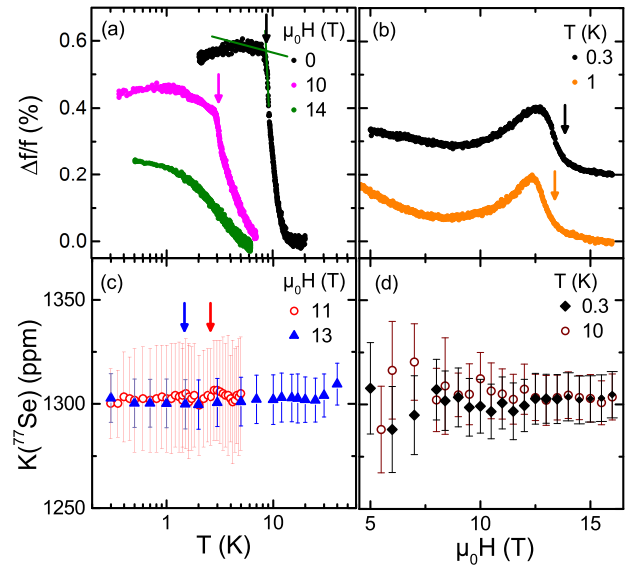}
	\caption{\label{fig:K}
		(a) Temperature-dependent detuning of the NMR tank circuit. The arrows
		denote $T_c$, determined from linear fits, as exemplified for the 0 T curve.
		(b) Field-dependent detuning, with the data at 0.3 K shifted vertically for
		clarity. The arrows denote fields where an order-disorder transition of the
		vortex lattice, manifested as peaked enhancement, is completed.
		(c) Temperature dependence of the $^{77}$Se Knight shift. The error bars are given
		by the spectral linewidth, the arrows denote $T_c$. (d) Field dependence of the
		$^{77}$Se Knight shift at 0.3 and 10 K. The magnetic field was always
		applied parallel to the $c$ axis.}
\end{figure}

\section{Results and Discussion}
The obtained results from the RF reflection measurements, shown in
Fig.\ \ref{fig:pd} as blue circles, are in very good agreement with
previously reported data for the boundary of the A phase \cite{kasaharapnas111}.
The characteristic temperatures and magnetic fields
that determine the boundary of the A phase were extracted as intersection
points of linear fits to the data above and below the respective slope
changes in the temperature- and field-dependent sweeps, see
Fig.\ \ref{fig:K}(a).
As shown in Fig.\ \ref{fig:K}(a), in the temperature-dependent sweeps at
fields corresponding to the A phase, the transition from the normal to the
superconducting state is manifested as a pronounced change of slope, and
we find $T_c =$ 8.7 K at 0 T. At 14 T, however, upon crossing the
transition to the B phase, no pronounced feature is observed. 

In the field-dependent sweeps at low temperatures (see Fig.\ \ref{fig:K}(b)), the resonance frequency at first monotonically decreases due to vortex formation in the Shubnikov
phase, and shows a maximum closely below the transition between the A
and B phase. This maximum is attributed to an order-disorder transition of the vortex lattice, as was observed also by magnetic-torque measurements of FeSe, see supplemental information of
\cite{kasaharapnas111}. 

Similar to the RF reflection experiments, a change of the RF volume
penetration at the transition to the superconducting state can be
probed via the intensity of the NMR spectra, see inset of Fig.\ \ref{fig:pd}.
At 11 T, the intensity decreases by about $10\%$ below $T_c = 2.6$ K
due to vortex formation, and drops abruptly at around 0.5 K. The latter
feature is attributed to a highly non-monotonic temperature dependence
of the surface resistance below $T_c$, as revealed by RF impedance
measurements \cite{liarxiv160505407}. In contrast, at 15 T, the
intensity decreases sharply at $T_c \simeq 0.7$ K and reaches less
than $1\%$ of the normal-state value at lowest temperature.
We note that the RF field amplitude is approximately several mT in the
NMR intensity and a few ten $\mu$T in the RF reflection experiments,
respectively. Apparently, the electronic properties of the B phase yield
no pronounced electrodynamic response when probed via a weak RF field,
but generate pronounced shielding effects when stimulated with a strong
RF field. The origin of this behavior is unclear, so far.

Having established the phase diagram, we next turn to our investigation
of the intrinsic electronic properties, probed via NMR observables.
In general, the NMR Knight shift is defined as $K = K_\text{orb} +
K_\text{s}$, being composed of a constant orbital part $K_\text{orb}$ and
a spin part $K_\text{s}$, defined as $K_\text{s} = A_{hf} \cdot \chi$,
with the hyperfine coupling $A_{hf}$ and the electronic dc susceptibility
$\chi$. From previous $^{77}$Se NMR results of FeSe above the nematic
transition temperature, a hyperfine coupling of $A_{hf} = 3.77$ T/$\mu_{B}$
was reported \cite{boehmerprl114}. This value is large compared to the
hyperfine couplings of other nuclear isotopes used in the study of
iron-based superconductors \cite{terasakijpsj78,kitagawajpsj77,
Kitagawajpsj78,baekepj85}. Hence, the $^{77}$Se Knight shift and linewidth
in FeSe are very sensitive to any static local-field contribution. We
determine the Knight shift from the first spectral moment of our data,
using a nuclear gyromagnetic ratio $\gamma(^{77}$Se$)/2 \pi = 8.13$ MHz/T.
The $^{77}$Se NMR spectra yield a typical linewidth (FWHM) of about 3 kHz,
in line with previously reported results \cite{baeknmat14,boehmerprl114}.
We take the FWHM as very conservative error for the Knight-shift data,
see Fig.\ \ref{fig:K}.

For precise determination of the $^{77}$Se Knight shift, the magnetic
field was repeatedly calibrated with a $^{63}$Cu NMR reference during
the experiments. For all fields, the temperature dependence of the Knight
shift shows no noticeable change below 30 K, see Fig.\ \ref{fig:K}(c).
It increases only for $T \geq$ 30 K, consistent with previous reports
\cite{baeknmat14,boehmerprl114}. In particular, the $^{77}$Se NMR spectra
recorded at 0.3 K, deep in the superconducting state, yield, within error
bars, no decrease of the shift within the whole field range of our
measurements, including the transition between A and B phase [Fig.\
\ref{fig:K}(d)]. This is quite surprising, since, in a spin-singlet
superconductor, a reduction of the local spin susceptibility, driven
by the formation of Cooper pairs, occurs when approaching zero field
at temperatures much smaller than $T_c$. Therefore, we conclude that
the observed Knight shift of about 1300 ppm is purely of orbital origin,
and that the static uniform electronic spin susceptibility in FeSe is
extremely small at temperatures approaching the superconducting state.

This finding is further corroborated by the very small linewidth and
purely Lorentzian NMR lineshape in the whole parameter range of our
study. Any finite contribution from local susceptibilities would, with
the large $^{77}$Se hyperfine coupling mentioned above, immediately lead
to Gaussian line broadening. Also, the linewidth of the Redfield pattern,
resulting from a periodic array of local orbital magnetization in the
vortex lattice, is estimated to be much smaller than the linewidth of
the $^{77}$Se spectra in the high-field regime of our study
\cite{Curro2009}. With the given hyperfine coupling and linewidth, we
estimate the upper limit of the uniform static susceptibility to about
$2 \times 10^{-5}$ emu/(G mol). Further, our findings confirm that the
splitting of the $^{77}$Se line, observed for fields applied along the
$ab$ planes and interpreted as a microscopic signature of nematic order
in FeSe, stems from the anisotropic orbital
polarization in the orbital-ordered domains \cite{baeknmat14, boehmerprl114}.

\begin{figure}[tbp]
	\centering
	\includegraphics[width=0.95\columnwidth]{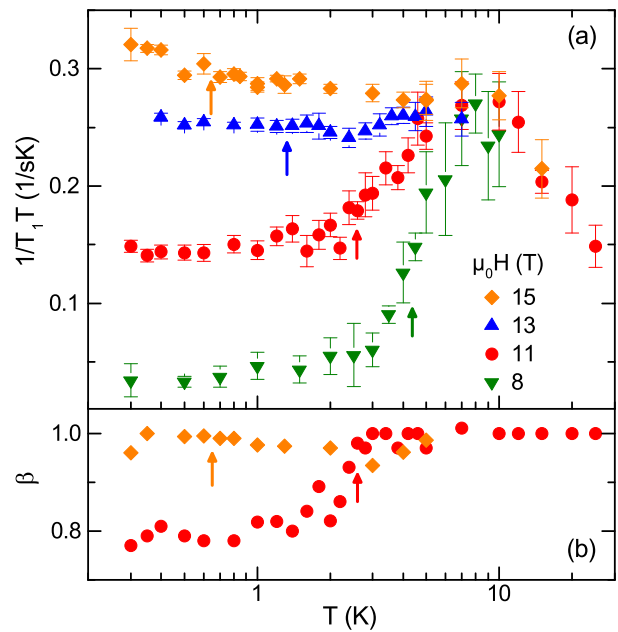}
	\caption{\label{fig:1T1T}(a) Temperature-dependent $1/T_{1}T$ of
FeSe for fields parallel to the $c$ axis. The arrows denote $T_c$ as
determined from the RF reflection and NMR intensity experiments,
respectively. (b) Stretching exponent $\beta$ of the $T_1$ relaxation
as a function of temperature.}
\end{figure}

Next, we turn to the discussion of the low-energy spin fluctuations.
As shown in Fig.\ \ref{fig:1T1T}(a), for $T >$ 10 K, $1/T_{1}T$ is only
weakly field dependent, in good agreement with previous results
\cite{baeknmat14,boehmerprl114,Kasahara2016,Shi2018}. For decreasing $T <$ 10 K
and $\mu_0H <$ 13 T, we find a decrease of $1/T_{1}T$, clearly starting
from well above $T_c$. These results are in line with recent reports of
pre-formed Cooper pairs and associated pseudogap behavior, leading to
a depletion of the density of states above the superconducting
condensation temperature \cite{Kasahara2016,Shi2018}.
The pseudogap sets in below about 10 K, whereas the spin
part of the Knight shift decreases already at higher temperatures(see Fig.\ \ref{fig:K}(c)),
finally becoming smaller than the spectroscopic linewidth below 30 K. Consequently, the decrease of the static spin susceptibility is
not related to the pseudogap formation.
	
Upon crossing the superconducting transition, $1/T_{1}T$ drops further - without showing any feature that might be associated with $T_c$ -
and levels off at about 2 K at 11 T and 3 K at 8 T, in good agreement
with the vortex liquid-solid transition indicated by the peak fields
obtained from the torque data reported by Kasahara et al. \cite{kasaharapnas111},
as well as our RF reflection measurements. At lower temperatures,
$1/T_{1}T$ is almost constant, and increases with magnetic field.
For higher magnetic fields, at 13 T, the relaxation rate does not
decrease below $T_c$, and finally, at 15 T, despite the transition to
superconductivity as probed by the abrupt decrease of the NMR signal
intensity (cf.\ inset of Fig.\ \ref{fig:pd}), $1/T_{1}T$ even
monotonically increases towards lowest temperatures. These
observations are in line with a field-driven suppression of the
pseudogap, being associated with preformed pairs of the superconducting
condensate, as predicted for the BCS-BEC crossover regime \cite{Kasahara2016,Randeria2014}.

In Fig.\ \ref{fig:1T1T}(b), the temperature dependence of the stretching
exponent $\beta$ is shown for 11 and 15 T. In the normal-conducting
state, we find $\beta \approx$ 1, indicating a single $T_1$ and
correspondingly a spatially homogeneous electronic state in the whole
sample volume. At 11 T, when entering the A phase, $\beta$ becomes
smaller than unity below $T_c$, in line with the presence of a vortex
lattice with superconducting regions and vortex cores with increased
quasiparticle density. In stark contrast, at 15 T, in the B phase,
$\beta$ remains close to unity at temperatures below $T_c$, evidencing
a spatially homogeneous electronic state.

\begin{figure}[tbp]
		\centering
	\includegraphics[width=0.95\columnwidth]{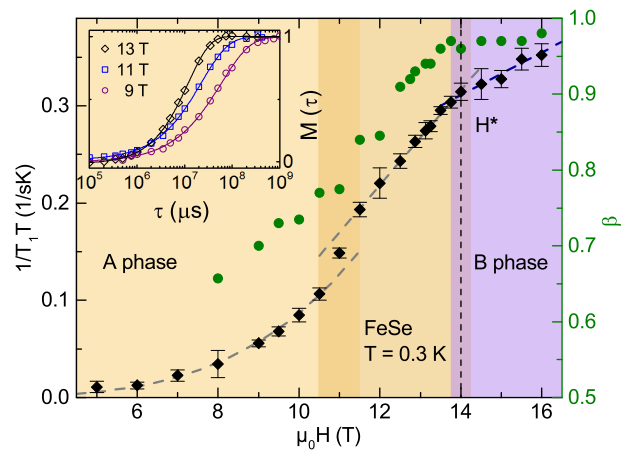}
	\caption{\label{fig:T1vsB} Magnetic-field dependence of $1/T_{1}T$
(diamonds) and the corresponding stretching exponent (circles) at 0.3 K,
deep in the superconducting state of FeSe. Dashed lines are guides to
the eye. The vertical dashed line denotes the transition field $H$* between
A and B phases. The inset shows selected nuclear relaxation curves, the
stretched behavior clearly decreases with increasing field.}
\end{figure}

The field dependence of $1/T_{1}T$ at 0.3 K (Fig.\ \ref{fig:T1vsB}),
deep in the superconducting state, is determined both by the structure
of the vortex lattice and the gradual suppression of the
superconducting gap amplitudes. Again, the presence of a vortex lattice
gives rise to a spatial variation of quasiparticle densities, yielding
a distribution of $T_1$ relaxation times with $\beta$ becoming smaller
than unity. Towards high fields, the superconducting gap amplitudes
decrease, and the vortex density with the corresponding volume fraction
of the normal-conducting vortex cores increases. In consequence, the
overall $T_1$ distribution sharpens, and $\beta$ approaches unity.

The field dependence of $1/T_{1}T$ yields two transitional regimes at
around 11 and 14 T, respectively. As was reported from measurements
of the thermal Hall coefficients, the transverse thermal conductivity
changes sign at around 12 T and 0.59 K, indicating that the
quasiparticles that determine the thermal conduction change from
electron-like to hole-like \cite{watashigearxiv1607}. The clear change
in the field dependence of $1/T_{1}T$ at around 11 T is likely driven
by the same phenomenon, namely, a field-driven closure of the smaller,
anisotropic gap on the electron pocket. Here, the field-dependent
increase of $\beta$ shows a monotonic variation of the spatial
dependence of the low-energy quasiparticle density.

The second change of the field dependence of $1/T_{1}T$ occurs around
$\mu_0 H^{\star} \approx$ 14 T, which gives evidence of a field-driven
transition to a distinct bulk superconducting state, in very good
agreement with features found in previous reports based on thermodynamic
quantities \cite{kasaharapnas111,watashigearxiv1607}. Above 14 T, in the
B phase, the field dependence of $1/T_{1}T$ becomes approximately linear
with a significantly smaller slope, without saturation up to 16 T. More importantly,
the stretching exponent $\beta$ saturates close to unity above 14 T,
evidencing spatially homogeneous low-energy quasiparticle excitations.
Since the magnetic-torque measurements show no anomaly at $\mu_0 H^{\star}$,
a Lifshitz transition or spin-density wave order as driving mechanism for
the transition to the B phase are very unlikely \cite{kasaharapnas111}.

Finally, we comment on the compatibility of our results with an FFLO
state underlying the B phase \cite{kasaharapnas111,watashigearxiv1607}.
The spectroscopic signature of a spatially inhomogeneous superconducting
state, as predicted by FFLO, with coupled modulated local susceptibility,
would be an inhomogeneous broadening of the spectral line. This was
observed in NMR studies of the FFLO states in the organic superconductors
$\kappa$-(ET)$_2$Cu(NCS)$_2$ \cite{WrightPRL2011} and $\beta^{\prime
\prime}$-(ET)$_2$SF$_5$CH$_2$CF$_2$SO$_3$ \cite{KoutroulakisPRL2016}.
However, as $K_s$ is smaller than the $^{77}$Se linewidth in FeSe, a
corresponding modulation would not be resolved. As to the dynamic
properties, a spatially inhomogeneous superconducting state would
give rise to a local variation of the quasiparticle densities and result in a stretched relaxation,
unless strong mechanisms of nuclear spin diffusion are at play. However, spatial inhomogeneities arising from an FFLO state are in contrast to our observations of $\beta = 1$ for the B phase. We, therefore, may exclude a simple
FFLO state as origin for the B phase.

\section{Summary}

In summary, RF volume penetration and $^{77}$Se NMR measurements
of the static and low-energy dynamic susceptibility of
single-crystalline FeSe at fields between 5 and 16 T and temperatures
down to 0.3 K reveal that the high-field B phase represents a distinct
bulk superconducting state of a highly spin-polarized Fermi liquid
with nonlinear RF response of the surface conductivity. The B phase
yields, within our experimental resolution, no spatial modulation of
either the density of low-energy quasiparticle excitations or the
static local susceptibility, as evidenced by a non-stretched
spin-lattice relaxation and the absence of a discernible inhomogeneous
line broadening, respectively. Rather, $1/T_{1}T$ increases monotonically
towards low temperatures and increasing fields in the B phase, which
sets it apart from standard BCS bulk superconductivity with gapped
excitations. Further, measurements of the orbital part of the NMR
Knight shift deep in the superconducting state reveal that the static
spin susceptibility in FeSe becomes extremely small below 30 K, despite
the presence of pronounced spin fluctuations. In line with previous
results, $1/T_{1}T$ reveals a gapped behavior of the low-energy spin
fluctuations already well above $T_c$ in a wide field range, indicating
pseudo-gap formation due to preformed Cooper pairs, which underlines
the unusual superconductivity of FeSe in the BCS-BEC crossover regime.

\begin{acknowledgments}
We thank Y. Matsuda, S.-L. Drechsler, V. Grinenko, S. Arsenijevi\'c, and S. Kasahara for valuable
discussions. We acknowledge support from the Deutsche Forschungsgemeinschaft
(DFG) through GRK 1621 and the W\"{u}rzburg-Dresden Cluster of Excellence on
Complexity and Topology in Quantum Matter--$ct.qmat$ (EXC 2147, Project
No.\ 390858490), as well as by the HLD at HZDR, a member of the European
Magnetic Field Laboratory (EMFL). Z.T.Z. was financially supported by the
National Natural Science Foundation of China (Grant No.\ 11304321) and by
the International Postdoctoral Exchange Fellowship Program 2013
(Grant No.\ 20130025).
\end{acknowledgments}

\end{document}